\documentclass[%
 aip,
 amsmath,amssymb,
 preprint,
]{revtex4-1}

\usepackage{graphicx}% Include figure files
\usepackage{dcolumn}% Align table columns on decimal point
\usepackage{bm}% bold math
\usepackage[utf8]{inputenc}
\usepackage[T1]{fontenc}
\usepackage{mathptmx}
\usepackage{etoolbox}
\usepackage[colorlinks=false, urlcolor=false,citecolor=false]{hyperref}
\usepackage[svgnames]{xcolor}

\makeatletter
\def\@email#1#2{%
 \endgroup
 \patchcmd{\titleblock@produce}
  {\frontmatter@RRAPformat}
  {\frontmatter@RRAPformat{\produce@RRAP{*#1\href{mailto:#2}{#2}}}\frontmatter@RRAPformat}
  {}{}
}%
\makeatother
\begin{document}

%\preprint{AIP/123-QED}
\title[]{Deterministic Transferable Planar Dielectric Mirrors for Investigating Strong Light–Matter Coupling}
% Force line breaks with \\
\author{Atanu Patra}
\altaffiliation{These authors contributed equally to this work}
 %\altaffiliation[]{}%Lines break automatically or can be forced with \\
\email{atanu.patra@uni-wuerzburg.de}

\author{Subhamoy Sahoo}%
\altaffiliation{These authors contributed equally to this work}
\email{subhamoy.sahoo@uni-wuerzburg.de}
\author{Johannes Düreth}%
\author{Simon Betzold}%
\author{Sven Höfling}

%\affiliation{Julius-Maximilians-Universität Würzburg, Physikalisches Institut, Lehrstuhl für Technische Physik, Am Hubland, 97074 Würzburg, Germany}%
\affiliation{Julius-Maximilians-Universität Würzburg, Physikalisches Institut and Würzburg-Dresden Cluster of Excellence ctd.qmat, Lehrstuhl für Technische Physik, Am Hubland, 97074, Würzburg, Germany}
% \homepage{http://www.Second.institution.edu/~Charlie.Author.}
%\affiliation{%
%Second institution and/or address%\\This line break forced% with \\
%}%

\date{\today}% It is always \today, today,

\begin{abstract}
Optical cavities play a central role in photonic and quantum technologies by enhancing light-matter interactions. In semiconductor microcavities, achieving high quality (Q)  factors typically relies on sophisticated epitaxial growth techniques, such as molecular beam epitaxy, which offer atomic-scale precision but are costly and limited in material compatibility. For dielectric microcavities, high Q factors can be achieved using dielectric Bragg mirrors. However, conventional deposition techniques for the top mirrors, such as plasma-enhanced chemical vapor deposition or sputtering, can damage embedded emitters. This limitation is particularly severe for van der Waals materials, especially atomically thin semiconductors. Moreover, the conventional top-mirror deposition can cover or degrade predefined metal contacts. Recovering electrical access typically requires additional lithography and etching steps. Here, a deterministic dry-transfer approach is developed to fabricate complete dielectric microcavities using both top and bottom SiO$_{2}$/TiO$_{2}$ Bragg mirrors without post-growth lift-off processes, reaching a Q factor $\sim$ 4$\times$10$^{3}$. 
%Here, a deterministic dry-transfer approach is developed to fabricate complete dielectric microcavities using both top and bottom SiO$_{2}$/TiO$_{2}$ Bragg mirrors, achieving IR-range Q $\sim$ 1198  without post-growth lift-off processes.  
Using a WS$_{2}$  monolayer as the active medium, clear signatures of strong exciton-photon coupling  are observed at both room temperature and cryogenic temperatures. These results demonstrate an efficient cavity fabrication approach that preserves the integrity of the emitter of layered materials, enabling 
next generation integrated photonic devices.

\end{abstract}

\maketitle

%\begin{quotation}
%The ``lead paragraph'' is encapsulated with the \LaTeX\ 
%\verb+quotation+ environment and is formatted as a single paragraph before the first section heading. 
%(The \verb+quotation+ environment reverts to its usual meaning after the first sectioning command.) 
%Note that numbered references are allowed in the lead paragraph.
%
%The lead paragraph will only be found in an article being prepared for the journal \textit{Chaos}.
%\end{quotation}

\section{\label{sec:level1}Introduction}

%... \\  
Van der Waals materials have opened up a new dimension in optoelectronic devices due to their mechanical flexibility, atomic-scale thickness, and, more importantly, the presence of stable excitonic resonances at room temperature, which are not accessible generally in conventional III-V or II-VI semiconductors\cite{ chernikov2014exciton, ugeda2014giant, he2014tightly, mak2016photonics, das2021transistors,de2025roadmap}. These materials become even more promising when integrated inside an optical microcavity, leading to the formation of exciton–polaritons \cite{liu2015strong, anton2021bosonic, lundt2019optical,liu2020nonlinear, waldherr2018observation, lamountain2021valley}. For monolayer transition metal dichalcogenides (TMDCs), Rabi splitting, which is a direct signature of strong light–matter interaction, typically lies in the range of 30 to 60 meV \cite{federolf2025embedding, zhao2023exciton, zhao2024room}, and can exceed 100 meV in plasmonic cavities \cite{bisht2018collective}. These high values allow the system to remain in the strong coupling regime under ambient conditions. This not only offers a new platform for advanced optoelectronic devices but also enables access to unique physical phenomena such as Bose-Einstein condensation, bound states in the continuum, and strong nonlinear optical responses \cite{anton2021bosonic, zhao2021ultralow, zhao2022nonlinear, weber2023intrinsic}. These systems can further enable a new generation of transport devices based on polariton propagation, as reported in previous studies 
\cite{wurdack2021motional, fitzgerald2025polariton, xie20252d, kaur2025polaritronics}.

%Apart from these unique physical properties, 2D materials have also attracted strong interest from the scientific community due to their cost effectiveness, particularly since the first successful demonstration of monolayer graphene using the scotch-tape exfoliation method. However, in order to observe exciton-polaritons, it is necessary to fabricate an optical microcavity. Depending on whether the cavity is based on semiconductor or dielectric materials, fabrication techniques such as molecular beam epitaxy, sputtering, plasma-enhanced chemical vapor deposition, or atomic layer deposition are commonly employed. These processes are typically time consuming, expensive, and require sophisticated infrastructure.

To observe exciton-polaritons, Fabry-Pérot cavities formed by top and bottom distributed Bragg reflectors (DBR) are one of the most versatile approaches for realizing exciton–polaritons\cite{liu2015strong, anton2021bosonic, lundt2019optical,liu2020nonlinear, waldherr2018observation, lamountain2021valley}. %Among the different cavity architectures, Fabry-Pérot cavities formed by top and bottom distributed Bragg reflectors (DBR) are one of the most versatile approaches for realizing exciton–polaritons\cite{liu2015strong, anton2021bosonic, lundt2019optical}.
However, in conventional fabrication, the DBRs are deposited over the entire substrate, which usually has lateral dimensions of several millimeters. In contrast, mechanically exfoliated 2D flakes rarely exceed lateral sizes of about 100 micrometers. As a result, more than 99 percent of the DBR area remains unused. This leads to material wastage and limits the miniaturization and integration potential of 2D polaritonic devices. This mismatch between the typical size of 2D flakes and the footprint of the cavity components calls for an advanced solution in which the dimensions of the DBRs are comparable to the flake size. In addition, direct deposition of the top DBR can degrade or damage the underlying 2D materials due to energetic particle bombardment. Although several studies have proposed transferable strategies to integrate top DBRs without damaging the active layers, these approaches often suffer from limited fabrication yield\cite{rupprecht2021micro}, rely on specific dielectric material combinations and substrates \cite{paik2023high, ge2024high}.\\
%Such an approach would allow multiple samples of different material systems to be placed close to each other on a single substrate, enabling parallel studies and compact device architectures.
In this work, we have demonstrated a method to fabricate micron-size optical cavities using transferable dielectric mirrors. The bottom and top DBRs are placed deterministically using the same dry transfer method commonly employed for 2D material assembly. DBRs with thickness up to $\sim$ 2.36 $\mu$m have been transferred. In the fabrication of DBRs, common dielectric materials like SiO$_{2}$ and TiO$_{2}$ are sputtered on a poly vinyl alcohol (PVA) coated glass substrate. We have observed strong coupling at room temperature, with Rabi splitting of $\sim$ 36 meV in monolayer WS$_{2}$. Strong coupling is maintained over a range of temperatures, enabling systematic tuning of exciton-cavity detuning. The resulting cavity exhibits a robust behaviour at low temperature and high vacuum ($\sim$10$^{-6}$ mbar).

%In this paper, we have demonstrated the method to fabricate micron-size cavity using transferable dielectric mirrors. The bottom and top DBRs were placed deterministically using the same dry transfer method commonly employed for 2D material assembly. DBRs with thickness up to $\sim$ 2 $\mu$m have been transferred. In this methods, common dielectric materials like SiO$_{2}$ and TiO$_{2}$ are sputtered on a \textcolor{blue}{poly vinyl alcohol (PVA)} coated glass substrate without using any special mirror or substrates. WS$_{2}$ monolayer was used to demonstrate the strong coupling at room temperature as well as lower temperature. The formed monolithic cavity showed a robust behaviour at low temperature and vacuum ($\sim$ 10$^{-6}$ mbar). This technique would be useful to study the strong coupling under optical excitation and especially for electrical excitation of field dependent studies. In the later case, the electrodes would be unaffected by placing the top DBRs deterministically without the requirement of the complex fabrication processes.

\section{\label{sec:level2}Results and Discussion}
\begin{figure}[h!]
	\centering
	\includegraphics[width=0.98\linewidth]{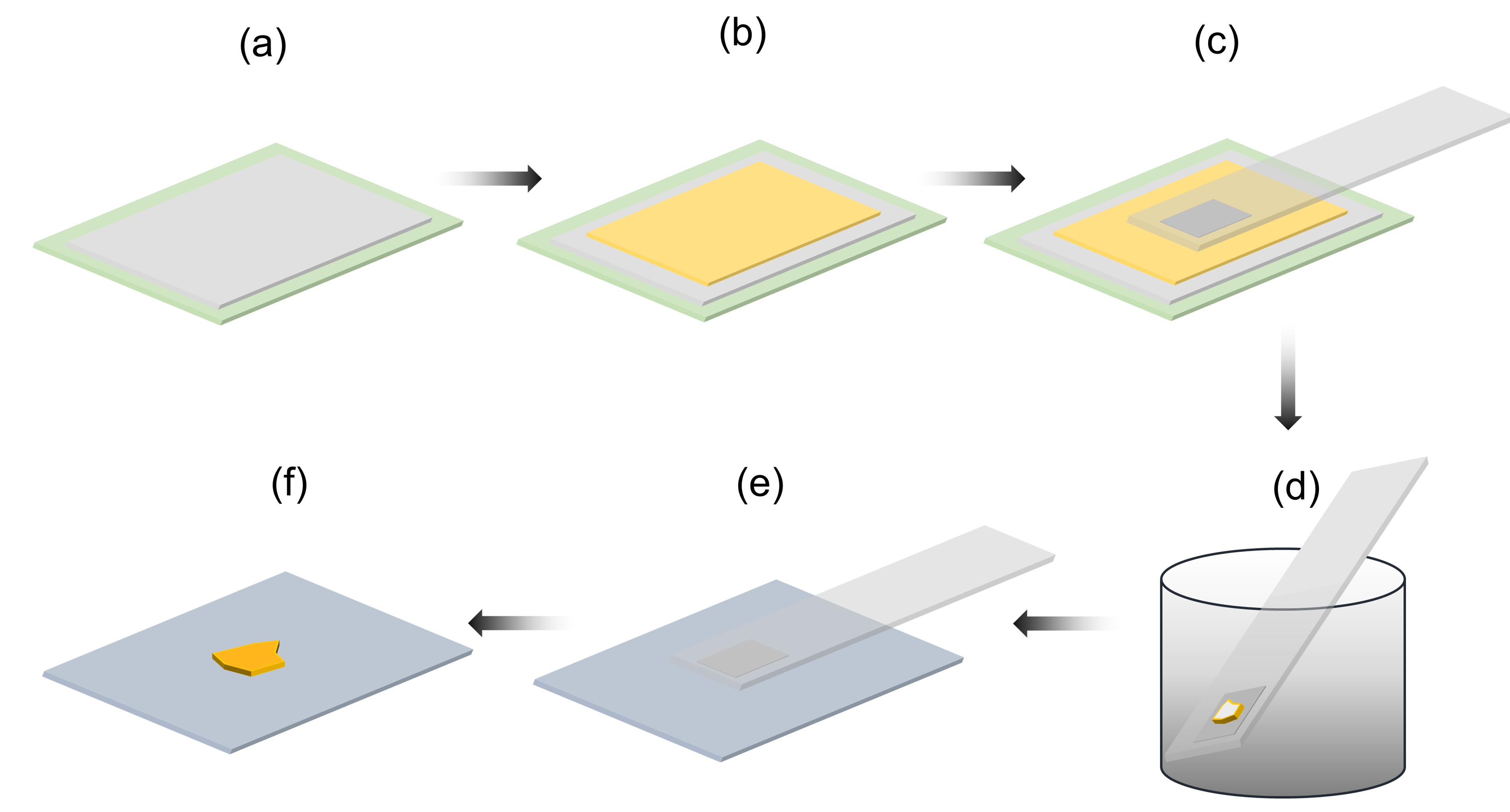}       
	\caption{The schematic diagram of the steps to fabricate the transferable DBRs is shown. The glass substrate (light green) with $\sim$100 nm PVA layer (grey) and sputtered DBRs (yellow) on PVA are shown in (a) and (b), respectively. The pickup of fragmented micro DBR using PDMS is shown in (c). The rinsing of the micro-DBR to remove PVA residue is shown in (d). The transfer process of fragmented DBRs from PDMS to Si substrate is shown in (e) and (f).   }
	\label{Fig:schematic_of_process_of_transferable_DBR}
\end{figure}

\begin{figure}[h!]
	\centering
	\includegraphics[width=0.98\linewidth]{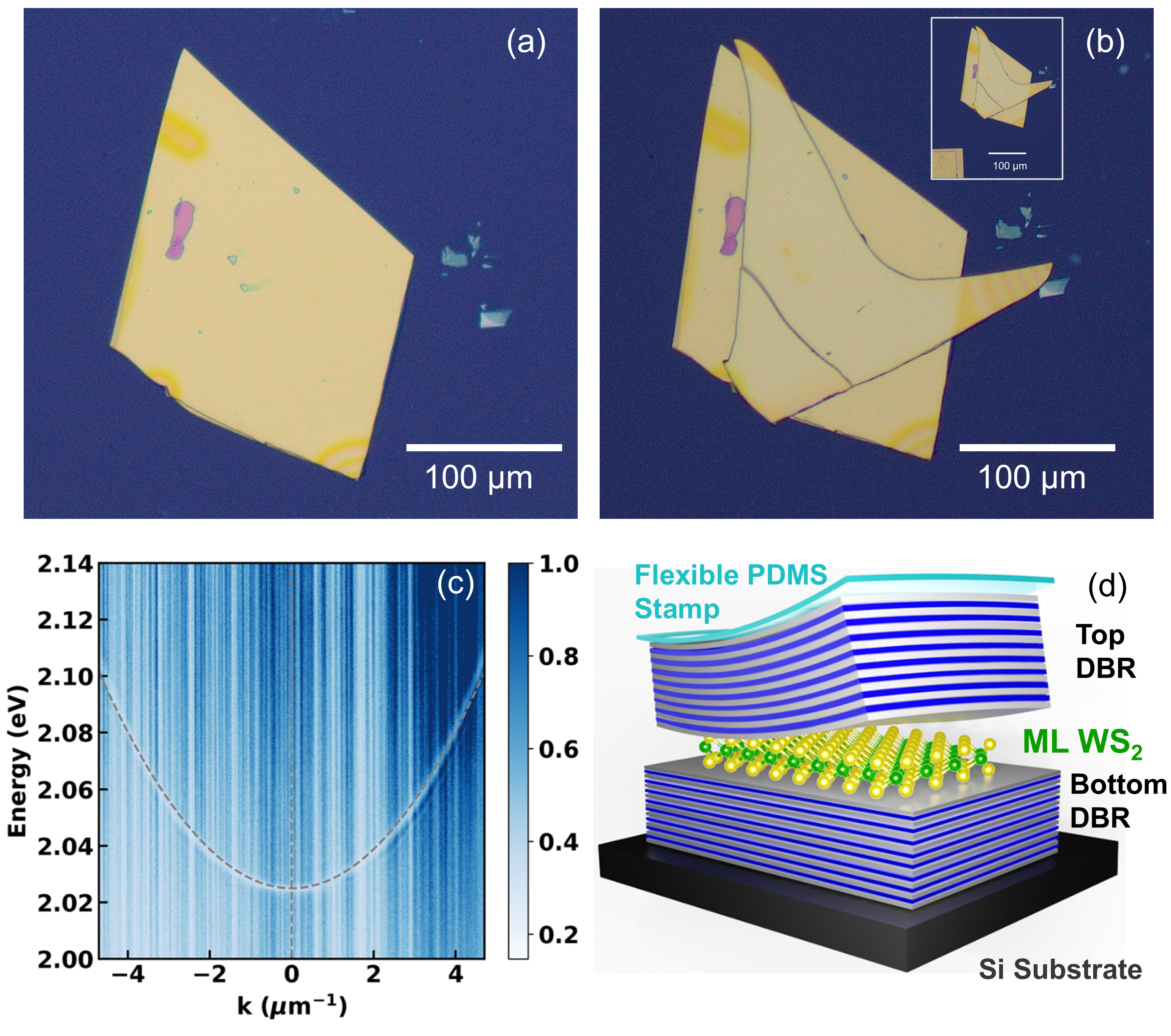}       
	\caption{(a) The transferred bottom DBR with the WS$_{2}$ monolayer flake and  (b) the full cavity structure with the top DBR on the Si substrate. The inset figure shows the full cavity structure with a gold marker. (c) The k-space reflection spectrum of the full cavity without monolayer WS$_2$. (d) The schematic diagram of the monolithic cavity structure with WS$_{2}$ monolayer. The SiO$_{2}$ and TiO$_{2}$ layers in both the top and bottom DBR are displayed by grey and blue colors. The embedded monolayer (ML) WS$_{2}$ is shown by the yellow-green colour.  }
	\label{Fig:Full_structure_and_cavity_mode}
\end{figure}

The schematic of the processes of developing transferable DBRs is shown in the Fig. \ref{Fig:schematic_of_process_of_transferable_DBR}. At first, the cleaned glass substrate was coated with aqueous solution of PVA using spin coating (Fig. \ref{Fig:schematic_of_process_of_transferable_DBR}.(a)). Then, the dielectric mirrors (SiO$_{2}$ and TiO$_{2}$) of thickness $\lambda$/4 were fabricated using radio frequency (RF) sputtering at a base pressure of $\sim$10$^{-4}$ mbar (Fig. \ref{Fig:schematic_of_process_of_transferable_DBR}.(b)). The top layer was ended with SiO$_{2}$ buffer layer of desired thickness to create a $\lambda$/2 cavity in the monolithic structure. Then the fraction of the DBRs was picked up using polydimethylsiloxane  (PDMS) (Fig. \ref{Fig:schematic_of_process_of_transferable_DBR}.(c)). The weak adhesive force between the SiO$_{2}$ and the PVA layer helped to break off the the DBRs while force was applied using PDMS. At this stage, the residue of the PVA may remain on DBRS. Since PVA is transparent in the visible and infrared (IR) region, this residue would not be a problem for optical measurement \cite{schnepf2017nanorattles}. However, it can hinder the dry transfer process. Moreover, these residues can behave as trap states for non-radiative recombination, thereby reducing the emission intensity \cite{kim2018photoluminescence}.
%However, it could create problem during the dry transfer process. 
To remove these residues, fragmented DBRs were rinsed with de-ionized (DI) water.
%Therefore, the fragmented DBRs were cleaned using de ionized (DI) water. 
The cleaned micro DBR was then transferred on a Si substrate (Fig. \ref{Fig:schematic_of_process_of_transferable_DBR}.(f)). This transferred micro DBR served as the bottom DBRs for the structure. Next,  using the same viscoelastic transfer process, the active material was placed on the bottom DBRs (see Fig. \ref{Fig:Full_structure_and_cavity_mode} (a)). To complete the microcavity, another fragment of the same  micro-DBR was flipped by transferring it on a second PDMS  after removing the residual PVA  (Fig. \ref{Fig:schematic_of_process_of_transferable_DBR}(d)). The inverted micro DBR, serving as the top DBR, was then transferred to the bottom structure to complete the full stack. 

The emission of the TMDCs spans from the visible to near-infrared (IR) region \cite{li2014measurement}. Consequently, the thickness of the dielectric pairs and total thicknesses of DBR stacks vary from small to large depending on the cavity mode and the number of mirror pairs, which determine the control of the transfer process. The increased stiffness of thicker flakes reduced their contact with the substrate during viscoelastic transfer, making precise placement challenging.
By using a micro DBR, we demonstrated deterministic transfer of the thick DBRs on a Si substrate. The approach enables the fabrication of DBRs that support the wide emission range of TMDCs.
%DBRs have cavity modes spanning from the visible to the infrared (IR) region. 
One such  cavity was designed for a mode at around 1.630 eV ( $\sim$ 760 nm), which could be useful for the systems emitting at near-IR (such as monolayer MoSe$_2$). This cavity consists of 10.5 pairs of SiO$_{2}$/ TiO$_{2}$ mirrors for both the bottom and top DBRs (inset of Fig. \ref{Fig:Q_value_MDS3}(a)). The thickness of each DRB is $\sim$ 2.36 $\mu$m. The cavity exhibits a quality factor (Q) of 3927, as shown in Supplementary Fig. \ref{Fig:Q_value_MDS3} (b).  For the emitters in the visible region, such as WS$_{2}$, a cavity (with a target cavity mode at 2.026 eV ($\sim$ 612 nm)) was fabricated using bottom and top DBRs with 8.5 pairs of $\lambda$/4 thick SiO$_{2}$/ TiO$_{2}$ mirrors. The total thickness of each fragmented DBRs is 1.47 $\mu$m. One such bottom DBR on a Si substrate is shown in Fig. \ref{Fig:Full_structure_and_cavity_mode}(a). The green patches on the DBR indicate the transferred WS$_{2}$ monolayer, along with a small bulk part. The complete fabricated cavity structure is shown in Fig. \ref{Fig:Full_structure_and_cavity_mode}(b). The DBRs were positioned in close proximity to a gold alignment marker, as illustrated in the inset of Fig. \ref{Fig:Full_structure_and_cavity_mode}(b).

The schematic of the full structure is displayed in Fig. \ref{Fig:Full_structure_and_cavity_mode}(d).  The white light reflectivity spectrum showed the cavity mode as evident in Fig. \ref{Fig:Full_structure_and_cavity_mode}(c). The quality factor of the cavity is determined from the full width at half maxima (FWHM) of the cavity mode. To obtain that, a vertical cut at k=0 (shown by the gray dotted straight line in Fig. \ref{Fig:Full_structure_and_cavity_mode}(c)) was taken, and the data was fitted after background correction, yielding a Q factor of 946. The theoretical Q factor of the cavity is 2076, as obtained from the transfer matrix (TMM) calculation. The cavity Q factor can be further enhanced by increasing the number of mirror pairs.

\begin{figure}[h!]
	\centering
	\includegraphics[width=0.98\linewidth]{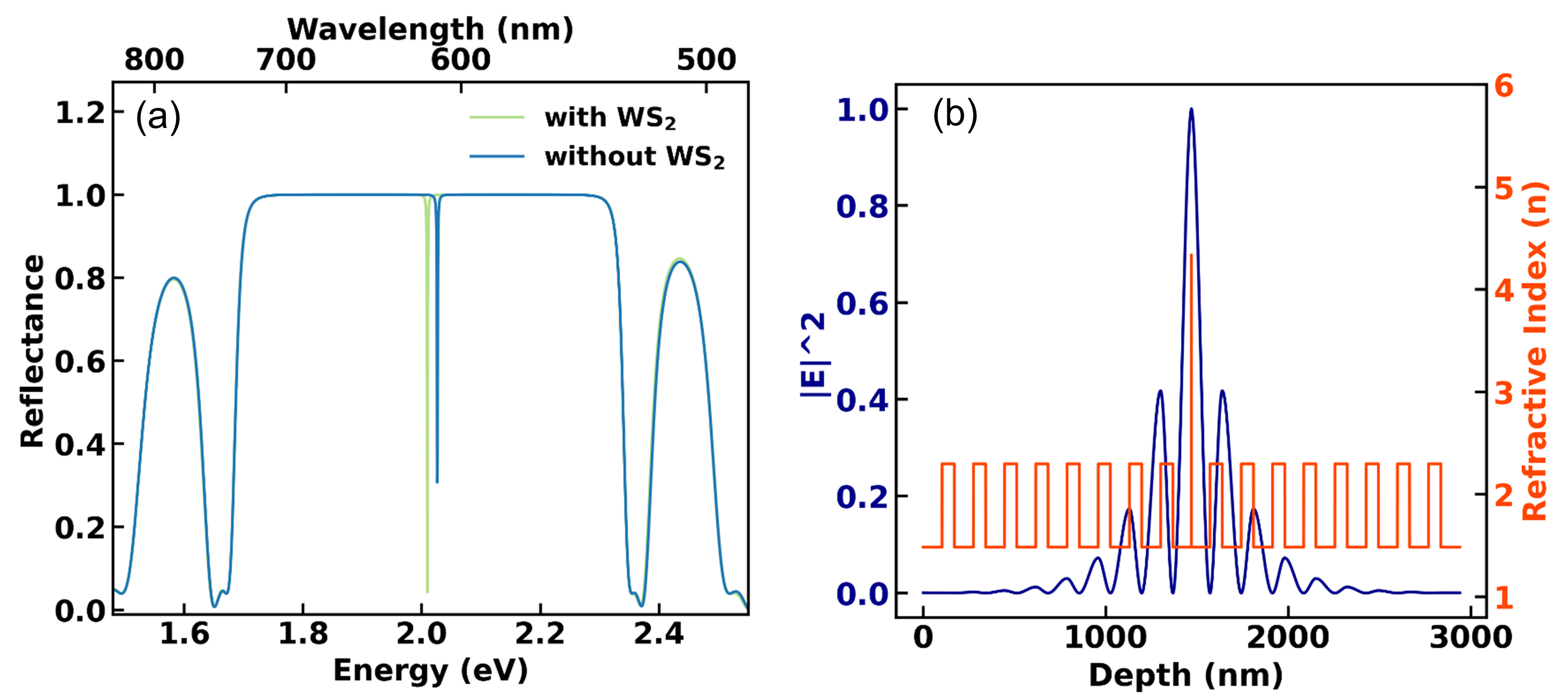}       
	\caption{(a) The variation of the cavity mode of the monolithic structure with (light green) and without (light blue) WS$_{2}$ flake as obtained from the transfer matrix calculation. (b) The distribution of the optical electric field (blue) at the cavity wavelength for the monolithic structure with embedded WS$_{2}$ monolayer is displayed as a function of the depth along with the refractive indices of the layers (red).}
	\label{Fig:Comparison_of_cavity_mode_TMM}
\end{figure}

\begin{figure*}[t]
	\centering
	\includegraphics[width=0.95\linewidth]{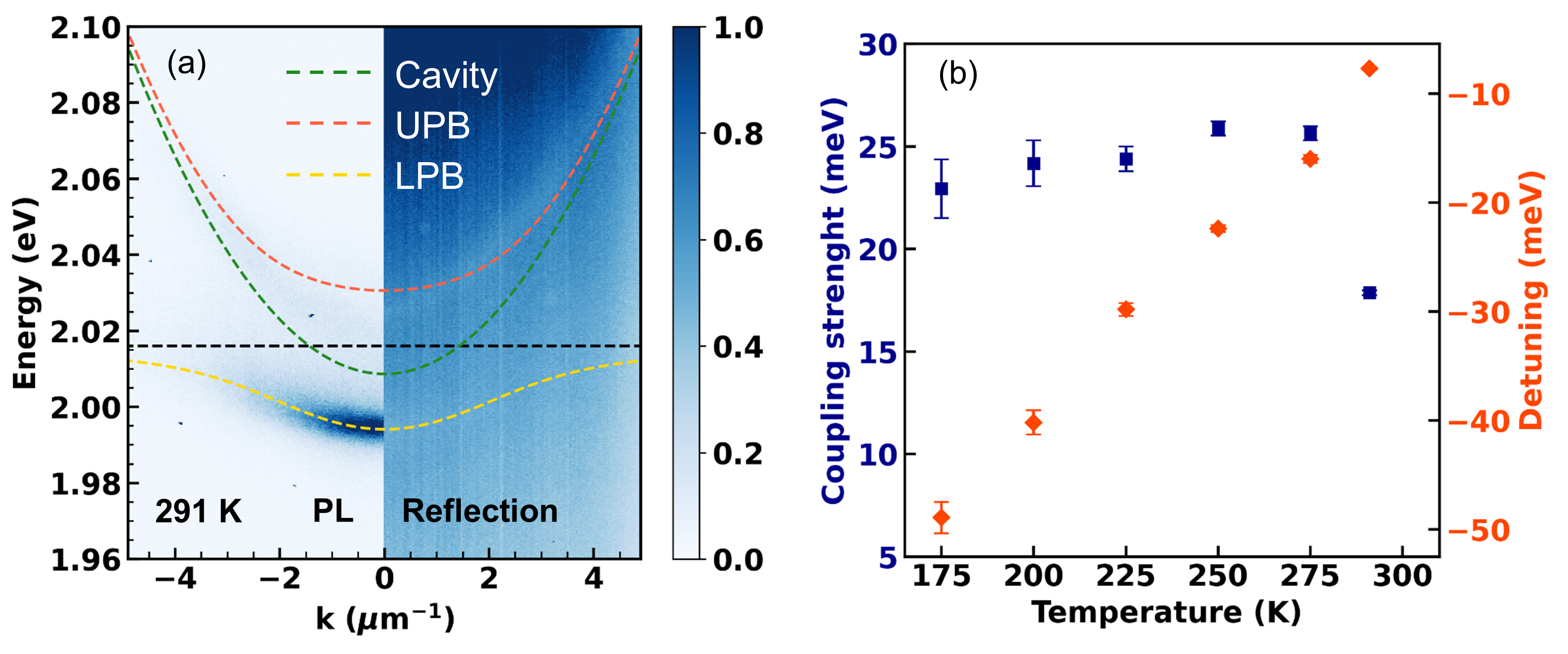}       
	\caption{(a) k-space PL (left side) and reflection (right side) of WS$_{2}$ monolayer at room temperature are shown. The exciton energy and the cavity mode are represented by broken black and green lines, respectively. The lower and upper polariton branches are shown by broken golden and red lines. (b) The variation of the coupling strength (blue square) with temperature is displayed along with the detuning (red diamond).}
	\label{Fig:k_space_PL_WS2_temperature_serise}
\end{figure*}

%To demonstrate the practicality of the microcavity and to study the strong coupling, WS$_{2}$ monolayer was used as emitter.
Using WS$_2$ as the emitter, we demonstrate the practical functionality of the microcavity and the onset of strong light-matter interaction. In the absence of WS$_{2}$ layer, the cavity mode was observed at 2.026 eV ($\sim$ 612.0 nm). With the incorporation of the monolayer WS$_{2}$ layer the cavity mode shifted to 2.010 eV ($\sim$ 617.0 nm) as observed from TMM simulation (Fig. \ref{Fig:Comparison_of_cavity_mode_TMM}(a)) for normal incidence. The refractive indices of the SiO$_{2}$ and TiO$_{2}$ layers were measured using ellipsometry, whereas the refractive index of WS$_{2}$ was taken from literature \cite{ermolaev2020spectral}. Since the bottom and top DBRs were acquired from the same mirror, the emitter would be at the maximum of the optical electric field. The simulated field distribution is shown in Fig. \ref{Fig:Comparison_of_cavity_mode_TMM}(b).

We measured the angle-resolved (k-space) photoluminescence (PL) under 2.331 eV (532 nm) (cw) excitation. The high-energy Bragg mode of the microcavity was at 2.369 eV ($\sim$ 523.3 nm) as obtained from the experiment and TMM simulation. However, the transmittance at 2.331 eV (532 nm) was $\sim$ 50\%. k-space PL was measured using a combination of an objective (Mitutoyo) of 0.65 numerical aperture (NA), a 4f lens configuration, and a monochromator (Andor Shamrock SR-500i) equipped with a charge-coupled device (CCD) (ANDOR Newton 971 EMCCD). The measurement was carried out at room temperature (without vacuum) as well as low temperature (at a pressure of $\sim$ 10$^{-6}$ mbar). The properties of the sample remained unaltered after cooling multiple time over a long period. It was also stable under vacuum, further corroborating the robustness of the structure.
%Therefore, the monolithic structure was robust. 

The k-space PL of the sample at room temperature is shown in the Fig. \ref{Fig:k_space_PL_WS2_temperature_serise}(a) (left side). A pronounced lower polariton branch (LPB) is observed compared to the upper polariton branch (UPB). The exciton energy was at 2.016 eV as obtained from the PL spectrum of the bare WS$_{2}$ flake. The k-space reflection spectrum is shown in the right half of the Fig. \ref{Fig:k_space_PL_WS2_temperature_serise}(a). The LPB is observed at the same energy as the PL spectrum. As the exciton energy varies with temperature, the energy difference between the cavity and the exciton, that is detuning, also varies accordingly. The exciton energy of a monolayer WS$_2$ varies by $\sim$ 80 meV as the temperature is reduced from room temperature to 4 K \cite{pei2022switching}. Therefore, the strong coupling was studied at various detunings by performing temperature-dependent measurements. The spectra were recorded from 175 K to 275 K at an interval of 25 K. %\textcolor{red}{\textit{The intensity of the emission linked to the bright exciton reduces with increasing temperature due to the shortening of the exciton lifetime (\textbf{add ref})}}. 
The spectra are shown in the supplementary Fig. \ref{Fig:k_space_PL_different_temperature}. The signature of UPB is not distinct in these spectra. However, a bare cavity signature was observed. In all the measurements at different temperatures, the signature of the bare cavity was present along with the polariton branches. The position of this was unchanged while LPB position varied. In case of the spectrum at room temperature (Fig. \ref{Fig:k_space_PL_WS2_temperature_serise}(a)), a pinhole was used to suppress the bare cavity contribution. Although the presence of the bare  bare cavity mode in the spectrum is not ideal, it enables the estimation of the effective refractive index of the cavity.

The spectra are fitted with coupled oscillator model to extract the coupling  strengths. In the analysis, only the interaction between exciton and cavity is considered. The Hamiltonian of the system is represented by a 2$\times$2 matrix, whose eigen values provide the energy of the polariton branches \cite{deng2010exciton}. The energies of the LPB and UPB are given by the equation \ref{equn:Energy_LPB_UPB}, where  $E_{c}$, $E_{x}$, and $g$ represent the energy of the cavity mode, exciton and coupling strength, respectively. 
\begin{equation}
	E_{U,L} (k) = \frac{1}{2} [E_{c} (k) +E_{x}] \pm \frac{1}{2} \sqrt{ [E_{c} (k) - E_{x}]^{2} + (2g)^{2}}
	\label{equn:Energy_LPB_UPB}
\end{equation} 
The Rabi energy ($\hbar \Omega_{R}$) is twice the coupling strength (2g). The detuning ($\delta$) is defined as the energy difference between the cavity mode and exciton ($E_{c} (0) - E_{x}$) at k=0, where k is the parallel component of the wave vector. In a realistic system, both the cavity mode and exciton exhibit finite broadening, which acts as a loss channel. The strong coupling persists in a system if 
\begin{equation}
	g > \sqrt{(\gamma_{c}^{2} + \gamma_{x}^{2})/2}
	\label{equn:strong_coupling_condition}
\end{equation} 
\noindent where  $\gamma_{c}$ and $\gamma_{x}$ are the half widths at half maximum of cavity mode and exciton energy \cite{keeling2020bose}.

The fits to the polariton branches using the coupled oscillator model are shown in Fig. \ref{Fig:k_space_PL_WS2_temperature_serise}(a). The fitted lower and upper polariton branch are indicated by dotted yellow and red lines, respectively. The dotted green parabola represents the cavity dispersion of the monolithic structure, including the emitter, while the broken black line indicates the exciton energy. The effective refractive index of the cavity is determined to be 1.65 which is consistent, as the refractive indices of the SiO$_{2}$ and TiO$_{2}$ are 1.48 and 2.29 near to the energy of cavity mode. At room temperature, the interaction strength is 17.9 meV. $\gamma_{x}$ of the exciton at room temperature is 12.0 meV, whereas the $\gamma_{c}$ of the cavity mode is 1.07 meV.  Therefore, condition for the strong coupling (equation \ref{equn:strong_coupling_condition}) is satisfied at room temperature. At 175 K, coupling strength is found to be 22.9 meV. As the temperature decreases, the coupling strength increases, as shown in Fig. \ref{Fig:k_space_PL_WS2_temperature_serise}(b). This observed temperature dependence can be attributed to the enhancement of the oscillator strength \cite{krustok2017local} along with a reduction in exciton dephasing and diminished non-radiative contribution to the linewidth \cite{moody2015intrinsic,selig2016excitonic,meshulam2025temperature}. This facilitates stronger and more pronounced light-matter interaction.

%This observed temperature dependence reflects the reduced exciton dephasing and diminished non-radiative contribution to the linewidth at lower temperatures \cite{moody2015intrinsic,selig2016excitonic,meshulam2025temperature}. This facilitates stronger and more pronounced light-matter interaction.

%This behavior can be attributed to the enhancement of the oscillator strength and reduction in the scattering process at low temperature (ref).

\section{\label{sec:level3}Conclusion}  
We have demonstrated a deterministic method to fabricate the monolithic structure using transferable mirrors composed of commonly used SiO$_{2}$ and TiO$_{2}$, for the TMDCs by a complete dry transfer process. The monolithic cavity containing WS$_{2}$ monolayer exhibited clear signature of strong coupling in ambient condition as well as at low temperature and high vacuum ($\sim$ 10$^{-6}$ mbar). The optical properties of the sample was unaltered after multiple round of cooling and vacuuming over a period of eight months, which showed the robustness of the sample. This approach would provide an avenue for studying the dynamics and nonlinear phenomena in many-body exciton-polariton interaction under optical excitation. Moreover, micro DBRs can be easily integrated with electrical platforms, paving the way towards electrically addressable photonic devices.  Additionally, this method provides a scalable, cost-effective, and material-efficient route for the realization of compact exciton-polariton devices based on van der Waals materials.   

%This approach would provide an avenue for studying the dynamics and nonlinear phenomena in many-body exciton-polariton interaction under optical excitation, especially in the presence of an electrical field.   Moreover, this method provides a scalable, cost-effective, and material-efficient route for the realization of compact exciton-polariton devices based on van der Waals materials.   

\section{\label{sec:level3}Acknowledgment}  
We gratefully acknowledge funding by the Deutsche Forschungsgemeinschaft (DFG, German Research Foundation) within the project HO 5194/16-1. S.H and A.P. acknowledge the funding from the lighthouse project IQ-Sense of the Bavarian State Ministry of Science and the Arts as part of the
Bavarian Quantum Initiative Munich Quantum Valley (15 02 TG 86). We acknowledge financial support from the Würzburg-Dresden Cluster of Excellence on Complexity, Topology, and Dynamics in Quantum Matter ctd.qmat (EXC 2147, DFG project ID 390858490).
We acknowledge Monika Emmerling for the fabrication of markers on the Si substrates.

\section{References}
\nocite{*}
\bibliography{aipsamp}% Produces the bibliography via BibTeX.

%\section{Supplementary}
\newpage

\renewcommand{\thefigure}{S\arabic{figure}}
\renewcommand{\thetable}{S\arabic{table}}
\setcounter{figure}{0}
\setcounter{table}{0}
\setcounter{equation}{0} 
\setcounter{enumi}{0} 
\setcounter{enumiv}{0} % For resetting the bibliography/citation numbering
\setcounter{page}{1}

\begin{figure*}[h!]
	\centering
	\includegraphics[width=0.8\linewidth]{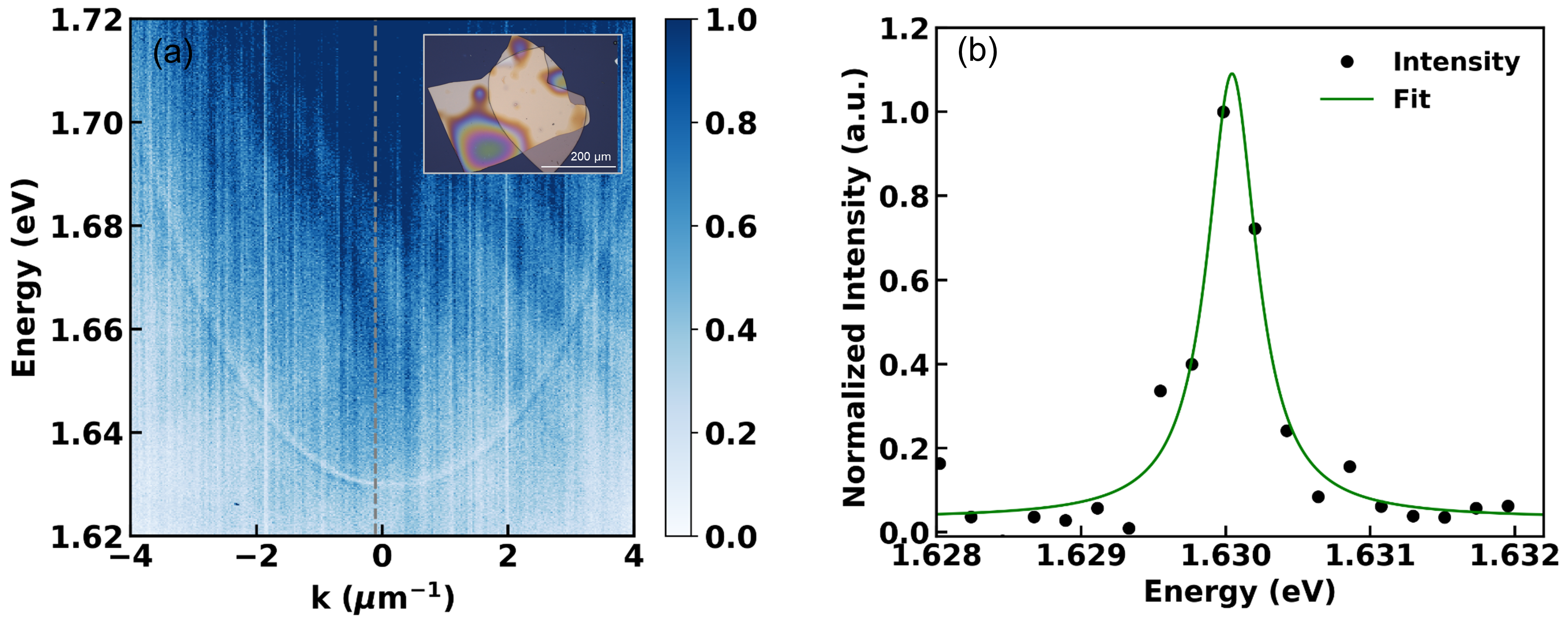}       
	\caption{(a) Dispersion of the cavity (cavity mode at 1.63 eV) is shown with the microscope image of the full structure in the inset. (b) The Lorentz fit of the cavity mode (line cut near k=0). The exprimental data points and the fitted curve are shown by black solid circles and solid green line, respectively.  }
	\label{Fig:Q_value_MDS3}
\end{figure*}

\begin{center}
	\textbf{\large{Supporting Information for}} \\
    \textbf{\large Deterministic Transferable Planar Dielectric Mirrors for Investigating Strong Light–Matter Coupling} \\

Atanu Patra$^1$, Subhamoy Sahoo$^1$, Johannes Düreth$^1$, Simon Betzold$^1$, Sven Höfling$^1$

\textit{$^1$Julius-Maximilians-Universität Würzburg, Physikalisches Institut and Würzburg-Dresden Cluster of Excellence ctd.qmat, Lehrstuhl für Technische Physik, Am Hubland, 97074, Würzburg, Germany}\\

\end{center}
The dispersion of the cavity with cavity mode at 1.630 eV is shown in Figure \ref{Fig:Q_value_MDS3}(a). A microscope image of the full structure is shown in the inset. The line cut near k=0 is plotted in Figure \ref{Fig:Q_value_MDS3}(b). The fit to the spectrum gives a Q factor of 3927. 

\begin{figure*}[h!]
	\centering
	\includegraphics[width=0.95\linewidth]{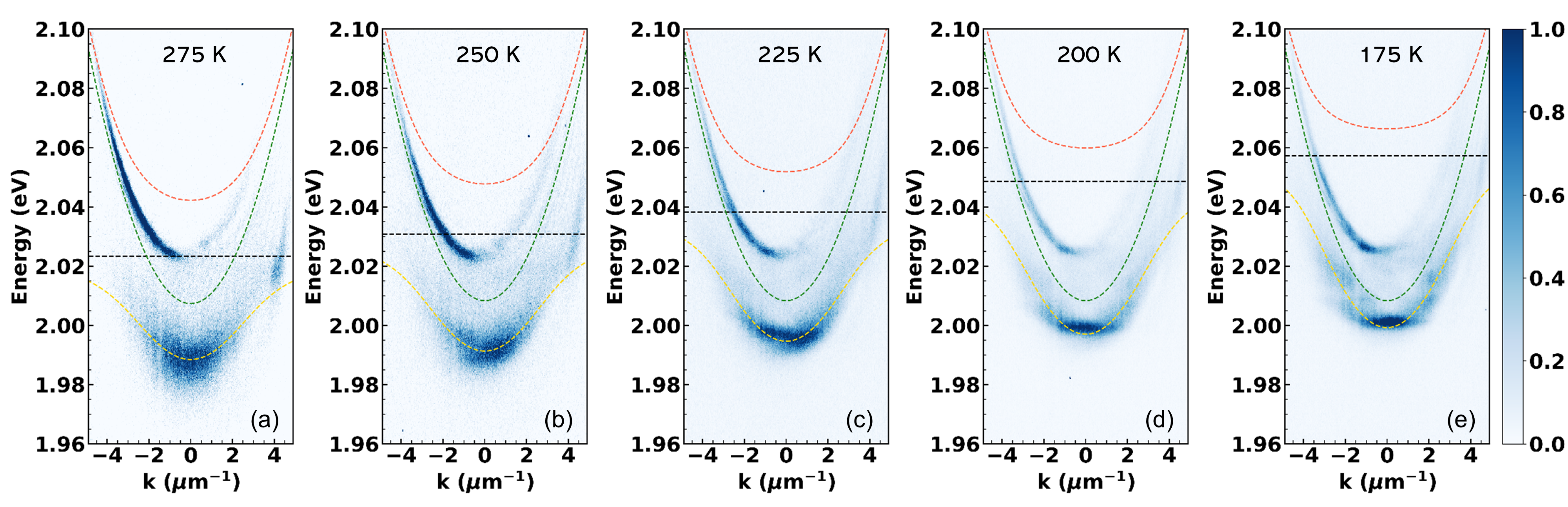}       
	\caption{ The variation of k space PL of  the WS$_{2}$
 monolayer at (a) 275 K, (b 250 K, (c) 225 K, (d) 200 K, and (e) 175 K are displayed  }
	\label{Fig:k_space_PL_different_temperature}
\end{figure*}

The variation of the k-space PL at different temperature of the WS$_{2}$
 monolayer are shown in the Figure \ref{Fig:k_space_PL_different_temperature}. The asymmetry observed in the cavity mode in the spectra arose due to the tilt in the sample.

\end{document}